\documentclass[twocolumn]{aastex63}
\pdfoutput=1 
\journalinfo{Accepted to the Astrophysical Journal}

\newcommand{\ha}{H$\alpha$\xspace}

\usepackage{nicefrac}

\usepackage{xspace}

\usepackage{ amssymb }
\usepackage{amsmath}

\newcommand{\angstrom}{\textup{\AA}\xspace}

\newcommand{\swift}{\textit{Swift}\xspace}

\newcommand{\nustar}{\textit{NuSTAR}\xspace}
\newcommand{\xte}{XTE~J1859+226\xspace}

\newcommand{\spm}[2]{\ensuremath{^{+#1}_{-#2}}}

\begin{document}

\title{An Optically-Discovered Outburst from XTE J1859+226} 

\author[0000-0001-8018-5348]{Eric C. Bellm}
\affiliation{DIRAC Institute, Department of Astronomy, University of Washington, 3910 15th Avenue NE, Seattle, WA 98195, USA}
\email{ecbellm@uw.edu}

\author[0000-0001-5538-0395]{Yuankun Wang}
\affiliation{DIRAC Institute, Department of Astronomy, University of Washington, 3910 15th Avenue NE, Seattle, WA 98195, USA}

\author[0000-0002-2626-2872]{Jan van Roestel}
\affiliation{Anton Pannekoek Institute for Astronomy, University of Amsterdam, 1090 GE Amsterdam, The Netherlands}

\author[0000-0001-6891-7091]{Rebecca A. Phillipson}
\affiliation{Villanova University, Department of Physics, Villanova, PA 19085, USA}

\author[0000-0002-8262-2924]{Michael W. Coughlin}
\affiliation{School of Physics and Astronomy, University of Minnesota, Minneapolis, Minnesota 55455, USA}

\author[0000-0001-5506-9855]{John A. Tomsick}
\affiliation{Space Sciences Laboratory, 7 Gauss Way, University of California, Berkeley, CA 94720-7450, USA }

\author[0000-0001-5668-3507]{Steven L. Groom}
\affiliation{IPAC, California Institute of Technology, 1200 E. California Blvd, Pasadena, CA 91125, USA}

\author[0000-0002-7718-7884]{Brian Healy}
\affiliation{School of Physics and Astronomy, University of Minnesota, 116 Church St SE, Minneapolis, MN 55455, USA}

\author{Josiah Purdum}
\affiliation{Caltech Optical Observatories, California Institute of Technology, 1200 E. California Blvd, Pasadena, CA 91125, USA}

\author[0000-0001-7648-4142]{Ben Rusholme}
\affiliation{IPAC, California Institute of Technology, 1200 E. California Blvd, Pasadena, CA 91125, USA}

\author{Peter Bealo}
\affiliation{American Association of Variable Star Observers, 185 Alewife Brook Parkway, Suite 410, Cambridge, MA 02138, USA}

\author{Stefano Lora}
\affiliation{American Association of Variable Star Observers, 185 Alewife Brook Parkway, Suite 410, Cambridge, MA 02138, USA}
\affiliation{Marana Space Explorer Center, Pasquali Road, Marana Di Crespadoro, Vicenza, 36070, Italy}

\author{Eddy Muyllaert}
\affiliation{American Association of Variable Star Observers, 185 Alewife Brook Parkway, Suite 410, Cambridge, MA 02138, USA}

\author{Ivo Peretto}
\affiliation{American Association of Variable Star Observers, 185 Alewife Brook Parkway, Suite 410, Cambridge, MA 02138, USA}
\affiliation{Marana Space Explorer Center, Pasquali Road, Marana Di Crespadoro, Vicenza, 36070, Italy}

\author{Erik J. Schwendeman}
\affiliation{American Association of Variable Star Observers, 185 Alewife Brook Parkway, Suite 410, Cambridge, MA 02138, USA}

\begin{abstract}

Using the Zwicky Transient Facility, in 2021 February we identified the first known outburst of the Black Hole X-ray Transient \xte since its discovery in 1999.
The outburst was visible at X-ray, UV, and optical wavelengths for less than 20 days, substantially shorter than its 320-day full outburst in 1999, and the observed peak luminosity was two orders of magnitude lower.
Its peak bolometric luminosity was only  $2\times 10^{35}$\,erg\,s$^{-1}$, implying an Eddington fraction of about $3\times10^{-4}$.
The source remained in the hard spectral state throughout the outburst. 
From optical spectroscopy measurements we estimate an outer disk radius of 10$^{11}$\,cm.
The low observed X-ray luminosity is not sufficient to irradiate the entire disk, but we observe a surprising exponential decline in the X-ray lightcurve.
These observations highlight the potential of optical and infrared (O/IR) synoptic surveys to discover low-luminosity activity from X-ray transients.

\end{abstract}

\section{Introduction}

Transient Black Hole X-ray Binaries (XRBs) exhibit large outbursts driven by instabilities in their accretion disks.
Canonical outbursts transition through a sequence of states of differing intensity, spectral hardness, and variability bracketed by a hard, low-luminosity state near quiescence and soft state at peak outburst flux \citep[e.g.,][]{Remillard:06:BHXRBs, Kalemci:22:BHSpectralTimingReview}.
These states reflect the changing conditions of the accretion disk, corona, and jet as the outburst progresses.
This evolution is broadly understood in the context of the Disk Instability Model \citep[DIM; for a review, see][]{Hameury:20:DIMReview}, in which increased densities cause the disk temperature to rise locally, ionizing hydrogen and creating a  viscous instability that propagates through the disk and increases the mass transfer onto the compact object.
In XRBs, irradiation of the disk by the compact object prolongs their outbursts and increases their recurrence times relative to dwarf novae \citep[e.g.,][]{King:98:SXTIrradiation}.
However, the exact drivers of the state transitions are still not understood in detail.

Some outbursts do not exhibit the full range of spectral states: such ``failed-transition'' or ``hard-only'' outbursts brighten without reaching the soft state \citep[e.g.,][and references therein]{Alabarta:21:FailedSXT}.
Almost 40\% of outbursts fail to transition, and individual binaries can exhibit both full and failed outbursts \citep{Tetarenko:16:WATCHDOG}.
Different mass accretion rates may influence whether state transitions occur.
\citet{Alabarta:21:FailedSXT} noted that in GX 339$-$4, the quiescent optical and infrared flux levels were higher prior to failed-transition outbursts when compared to successful ones.

Further observations of failed-transition outbursts can help pinpoint their causes.
As failed-transition outbursts are less luminous than canonical outbursts, they are more difficult to discover with all-sky X-ray monitors.  
Due to their faintness, some may fall in the phenomenological class of low-luminosity ($L_\mathrm{X} \sim 10^{34}$--$10^{36}$\,erg\,s$^{-1}$) Very Faint X-ray Transients \citep[VFXTs;][]{Wijnands:06:VFXT, Heinke:15:VFXRBlcs}, which may include outbursts from short-period accreting systems with intrinsically small accretion disks as well as failed-transition outbursts from portions of larger disks.
VFXTs are typically discovered through deep, cadenced observing programs by narrow-field X-ray telescopes \citep[e.g.,][]{Swank:01:RXTEBulge, Kuulkers:07:INTEGRALBulge, Bahramian:21:SwiftBulgeSurvey}, and so thus must sacrifice areal coverage for sensitivity.
Alternatively, optical and infrared observations by synoptic surveys or dedicated monitoring programs \citep[e.g.,][]{Zhang:19:SwiftJ1753MiniOutburst, Saikia:23:SwiftJ1910Optical} can also remove the selection effect imposed by X-ray monitor detection and provide a more comprehensive view of black hole accretion.

\xte was first discovered in outburst at 250\,mCrab in 1999 by the All-Sky Monitor on the Rossi X-ray Timing Explorer \citep{Wood:99:XTEJ1859Discovery} and peaked at 1.5\,Crab eight days later \citep{Focke:00:XTE}.
Followup observations revealed a 15th\,mag optical counterpart with broad Balmer and He II emission lines \citep{Garnavich:99:XTEJ1859Opt,Wagner:99:XTEJ1859}.
A radio counterpart with flux $\sim$10\,mJy was also detected \citep{Pooley:99:XTEJ1859Radio}.
X-ray observations revealed a hard power-law spectrum with an evolving Quasi-Periodic Oscillation \citep{Markwardt:99:XTEJ1859Xray,dalFiume:99:XTEJ1859QPO}. 
In combination these features suggested the discovery of a new Black Hole X-ray Transient.

In total the 1999 outburst lasted for about 320 days, with several late-time reflares peaking around $m_R \sim 15$\,mag, before returning to a quiescent magnitude of $m_R \sim 22.5$ \citep[e.g.,][]{Zurita:02:XTEJ1859Optical}.
The initial hard-state behavior of the outburst during the rise to a high-luminosity soft state provoked a recognition that this behavior is common among XRBs \citep{Brocksopp:02:XTEJ1859}.
\citet{Tomsick:03:XTEQuiescence} reported \textit{Chandra} X-ray observations of \xte in quiescence; its faint spectrum was consistent with an absorbed powerlaw  with a 0.3--8\,keV luminosity of 2.2$\times 10^{31}$ ($d$/8\,kpc)$^2$ erg\,s$^{-1}$.
Radial velocity observations by \citet{Filippenko:01:XTEJ1859RV} suggested the presence of a very massive black hole; however, later observations by \citet{Corral-Santana:11:XTEBH} provided a revised 
orbital period of 6.58\,hr, a
mass function of $4.5\pm0.6\,M_\odot$, and a lower limit $M_\mathrm{BH} > 5.42\,M_\odot$ with inclination $i < 70^\circ$.
As discussed in \citet{Tetarenko:16:WATCHDOG}, literature distance estimates for this source are sensitive to changing assumptions of the orbital period and secondary spectral type.
They adopt a fiducial 8$\pm$3\,kpc distance estimate taken from \citet{Hynes:05:XRBOIRSEDs}.  
We will use the same value in this paper, although we note that the shorter orbital period and later spectral type proposed by \citet{Corral-Santana:11:XTEBH} lead to model fits at a greater distance of $\sim$14\,kpc.

\citet{CorralSantana:10:XTERebrigthening} reported that in 2010 August \xte had re-brightened by $\sim$1\,mag in the optical and showed short-timescale flaring.
This brightening was not accompanied by an increase in the X-ray flux.

In 2021 February we identified a larger optical brightening \citep{2021ATel14372....1B} of \xte to $m_r \sim 18.9$\,mag using the Zwicky Transient Facility \citep[ZTF;][]{Bellm:19:ZTFOverview,Graham:19:ZTFScience}.
This was accompanied by an X-ray brightening  to a 0.3--10 keV flux of 7.9\spm{2.0}{1.5} $\times$ 10$^{-12}$ erg\,cm$^{-2}$\,s$^{-1}$ \citep{2021ATel14375....1B}.
However, continued monitoring by these facilities over the subsequent weeks saw the source flux peak and decline by 1.5\,mag.
This decline was also reported by the XB-NEWS project \citep{2021ATel14415....1C}.
Despite its low luminosity and short duration, in the taxonomy of \citet{Zhang:19:SwiftJ1753MiniOutburst}, the 2021 outburst is classified as a new outburst since the time elapsed since the full 1999 outburst is much larger than the duration of the 1999 outburst. 

In \S \ref{sec:observations} we describe the outburst discovery and observations.  
We analyze these observations in \S \ref{sec:analysis}.
We conclude in \S \ref{sec:discussion} with implications for future observations of low-luminosity outbursts.

\section{Observations} \label{sec:observations}

\subsection{Discovery of the Outburst} 

The Zwicky Transient Facility \citep{Bellm:19:ZTFOverview,Graham:19:ZTFScience} uses a large mosaic camera \citep{Dekany:20:ZTFObservingSystem} to survey the Northern Hemisphere sky ($\delta > -30^\circ$) in three optical bands ($g$, $r$, and $i$) to typical depths of 20.5\,mag with a cadence of two nights or faster \citep{Bellm:19:ZTFScheduler}.
Near-real time difference imaging pipelines \citep{Masci:19:ZTFDataSystem} identify transients, variables, and moving objects.
Motivated by a desire to identify compact binary outbursts and state changes \citep[cf.][]{Russell:19:OpticalPrecursorsXRBs}, we are monitoring the public alert stream \citep{Patterson:19:ZTFAlertDistribution} for a watchlist of known X-ray binaries using ANTARES \citep{Matheson:21:ANTARES}.

On 2021 February 4 ZTF observed a field containing \xte five times in $r$ band as part of its public twilight survey.
We received a notification from the ANTARES system of a new source, internally designated ZTF21aagyzqr, coincident with \xte\footnote{The ANTARES page for this object can be found at \url{https://antares.noirlab.edu/loci/ANT2021dn4jk}.}.
The ZTF detections were at $m_r \sim 18.9$\,mag, substantially brighter than the quiescent magnitude of  $m_R \sim 22.5$ \citep{Zurita:02:XTEJ1859Optical}.
We issued a circular \citep{2021ATel14372....1B} encouraging further observations in anticipation of additional brightening.
We also triggered \swift TOO observations and confirmed that the X-ray flux had increased by three orders of magnitude relative to the quiescent level \citep{2021ATel14375....1B}.
We used a customized Skyportal \citep{vanderWalt2019,Coughlin:23:Skyportal} instance (``Fritz'') for managing followup data.

\subsection{Optical}

\subsubsection{Photometry}

We obtained PSF forced photometry measurements on ZTF difference images \citep{Masci:19:ZTFDataSystem}.
We corrected the resulting differential photometry for the flux of the counterpart in the reference image, which is detected at $m_r \sim 21.7$\,mag in the $r$-band ZTF reference image.
The image is uncrowded in both the direct and difference images; the nearest PanSTARRS1 source is $1\rlap{.}^{\prime\prime}8$ arcseconds away and subtracts cleanly.
We excluded observations with \texttt{procstatus} values other than 0 or 57 to avoid biased photometry.
This returned more than 1400 forced flux measurements beginning in 2018 March.
Most were consistent with nondetections, but 214 were detections with signal-to-noise ratios greater than 3.
Of these detections, 124 were taken in a period of intensive monitoring on 2018 August 7--8.
The median $5\sigma$ upper limit was 20.9\,mag in $g$ band and 20.5\,mag in $r$ band.

We obtained difference image forced photometry from the ATLAS forced photometry service \citep{Tonry:18:ATLAS}.
We required reduced $\chi^2<20$, \texttt{err}$=0$, and magnitudes greater than 10 to reject clearly spurious measurements.
This yielded 1606 forced flux measurements beginning in 2015 October, of which 56 were detections with SNR$>3$.
The median $5\sigma$ upper limit was 19.2\,mag.

We imaged \xte in SDSS $g$, $r$, and $i$ bands with the Rainbow Camera of the Spectral Energy Distribution Machine \citep[SEDM;][]{Blagorodnova:18:SEDM} on the Palomar 60-inch telescope (P60) on 2021 February 8.
Automated reductions were performed using the methods described in \citet{Fremling:16:FPipe} and \citet{Blagorodnova:18:SEDM}.

In anticipation of further brightening of the source, we sent an Alert Notice\footnote{\#729; \url{https://www.aavso.org/aavso-alert-notice-729}} to the American Association of Variable Star Observers (AAVSO).
We retrieved the resulting photometry from the online download portal \citep{AAVSO}.

\subsubsection{Spectroscopy}

Along with three-color photometry, on 2021 February 8 we obtained a low-resolution spectrum of \xte using the P60 SEDM \citep{Blagorodnova:18:SEDM}.
The on-source exposure time was 2700\,s.
Automated reductions were performed using \texttt{pysedm} \citep{Rigault:19:pysedm, Kim:22:pySEDM}.

We obtained a longslit spectrum of \xte using the Low-Resolution Imaging Spectrometer \citep[LRIS;][]{Oke:95:LRIS} on  Keck I on 2021 February 15. 
The instrument was configured with a 1\farcs5 wide slit, the 560 dichroic, the 400/3400 grism, and the 400/8500 grating with a central wavelength of 7828\,\angstrom.
The blue-side exposure time was 500\,s and the red-side exposure time was 800\,s.
We reduced the data using \texttt{LPipe} \citep{Perley:19:lpipe}.

\subsection{X-ray} \label{sec:xrayobs}

We obtained a series of \swift observations between 2021 February 6--28 (Table \ref{tab:swift_obsids}). 
We used the \swift-XRT data products generator\footnote{https://www.swift.ac.uk/user\_objects/} \citep{Evans:07:XRTRepo, Evans:09:XRTAutoReduction} to produce count rates, hardness ratios, and spectra for each observation.
Due to low count rates we grouped observations 00031827006, 00031827007, and 00031827009 into a single summed spectrum.
After fitting a spectral model to this summed spectrum, we froze the spectral parameters and fit the flux normalization in each constituent spectrum.
Observations 00031827010, 00031827011, 00031827012, and 00031827013 had too few counts to produce spectra.

\begin{table*}[]
    \centering
    \begin{tabular}{|l|c|r|r|}
    \hline
    Obsid & Start time (MJD) & XRT Exposure (s) & UVOT filter \\
    \hline
00031827002 &  59251.97437910605 &  1990 & V \\
00031827003 &  59252.89778221208 &  991 & V \\
00031827004 &  59255.55894562874 &  847 & UVW1 \\
00031827005 &  59259.73500215374 &  934 & UVW1 \\
00031827006 &  59261.65729843129 &  869 & UVW2 \\
00031827007 &  59261.98584341670 &  742 & V \\
00031827009 &  59263.64414707874 &  864 & V \\
00031827010 &  59265.23678571347 &  899 & V \\
00031827011 &  59266.16805541601 &  719 & UVM2 \\
00031827012 &  59267.75982711509 &  1009 & V \\
00031827013 &  59273.22360766786 &  914 & UVW2 \\
\hline
    \end{tabular}
    \caption{\swift observations used in this work.}
    \label{tab:swift_obsids}
\end{table*}

\nustar observed \xte for 40.4\,ksec on 2021 February 13. 
We obtained publicly-available \nustar observations (obsid 90701305002, P.I.\ Harrison; see also \citealp{Draghis:21:XTENustarATel, Draghis:23:BHSpins}) from the HEASARC and processed them using standard procedures with \texttt{nuproducts} under Heasoft v6.28.
We extracted spectra for each module using 40\arcsec\ radius circular apertures at the position of \xte and extracted background spectra from 115\arcsec\  radius circular apertures placed on the same chip.

We performed spectral fits using ISIS \citep{Houck:00:isis}.
We fit the \swift-XRT data from 0.3-10\,keV and \nustar data from 3--30\,keV.
We used the \texttt{tbabs} absorption model with \texttt{vern} cross-sections \citep{Verner:96:vernCrossSections} and  \texttt{wilm} abundances \citep{Wilms:00:wilmAbundances}.
We rebinned the spectra using the binning scheme of \citet{Kaastra:16:OptimalBinningXray} and required a minimum of three counts per bin.
We fit using the \texttt{wstat} fit statistic \citep{Wachter:79:MaxLikeParameterEstimation, Arnaud:00:wstat}, which provides a maximum likelihood parameter estimate in the case of low-count Poisson-distributed source and background data.
We report uncertainties on the best-fit parameters at the $1\sigma$ confidence level.

We also retrieved publicly-available daily count rates for \xte from \textit{MAXI} \citep{Matsuoka:09:MAXI} and \swift-BAT \citep{Krimm:13:BatTransientMonitor}.

\subsection{UV}

We reduced the UVOT data for the \swift observations described in \S \ref{sec:xrayobs} using standard procedures with \texttt{uvotsource} under Heasoft v6.28. 
We used the standard 5\arcsec\ radius circular source extraction region and placed a 12\arcsec\ radius circular background region nearby.
No contaminating sources were apparent in the images.
We report source measurements with signal-to-noise ratio greater than 3; when undetected we provide 5$\sigma$ upper limits.

\section{Analysis} \label{sec:analysis}

\subsection{Outburst Duration} \label{sec:lightcurve}

\begin{figure*}
	\includegraphics[width=\textwidth]{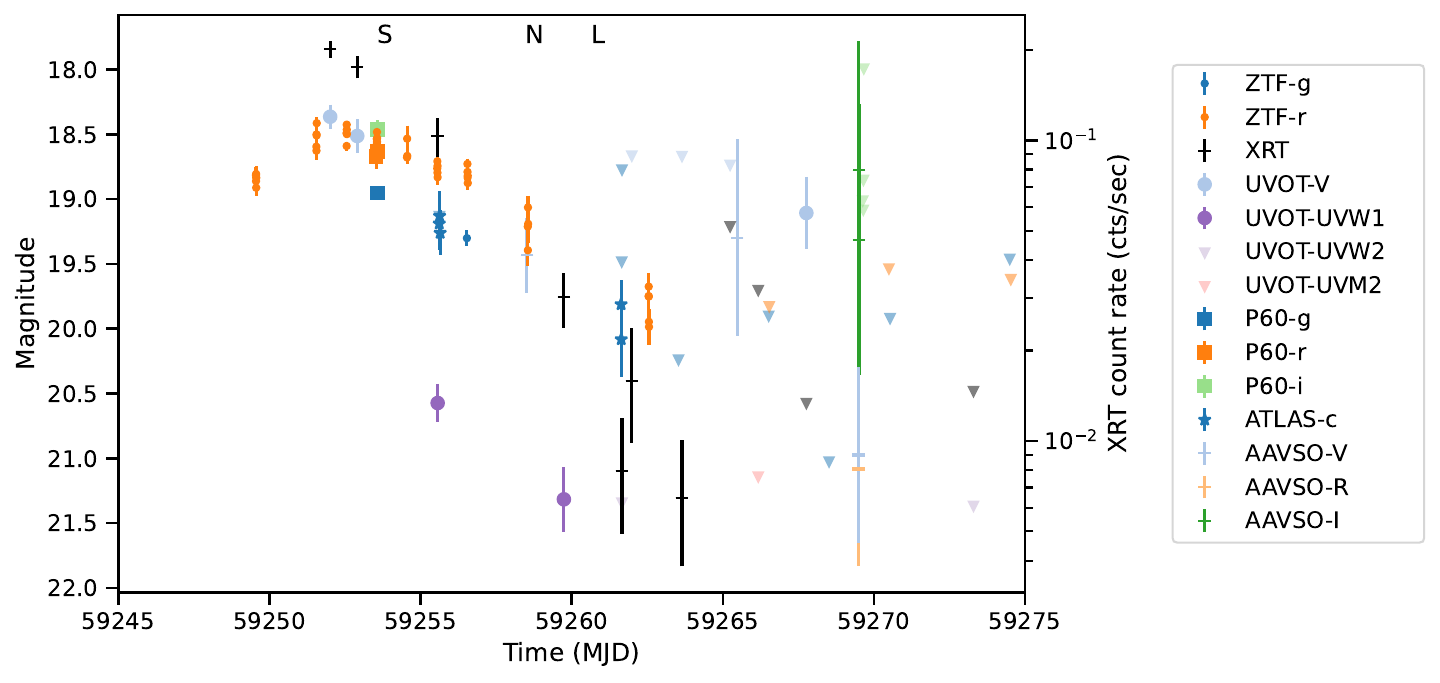}
\caption{Outburst lightcurve for \xte from ZTF, P60, AAVSO, and \swift-UVOT and XRT.
Detections (SNR$>$3) are marked with points; 5$\sigma$ upper limits are denoted with triangles.
Text labels indicate the times of SEDM spectroscopy (S), \nustar observations (N), and Keck-LRIS spectroscopy (L).
\label{fig:outburst_lc}}
\end{figure*}

Figure \ref{fig:outburst_lc} presents the optical and X-ray photometry of the outburst.
Because \xte was emerging from behind the sun, we do not have strong constraints on the onset of the outburst.
The last ZTF nondetection was on 2020 December 21 (\S \ref{sec:quiescent_lc}).
While we reported in \citet{2021ATel14372....1B} that ZTF had not detected \xte one night prior to the outburst onset, those observations were in a different, overlapping field and the position of \xte fell in a chip gap.
The upper limits included in the ZTF alert were estimates over the CCD rather than forced photometry at the source position.
Accordingly the ZTF observations on 2021 February 3 do not provide information on the state of \xte at that time.

Our \swift-XRT observations did not capture the rising phase of the outburst---the peak flux was observed in the first XRT observation 59 hours after the first ZTF detection.
In many cases the optical outburst is expected to precede an X-ray brightening \citep{Russell:19:OpticalPrecursorsXRBs}, so it is plausible that our first \swift observations are near the peak of the outburst in the X-ray.
Nevertheless, reflares and other temporal irregularities are common in XRB outbursts.
We searched other instruments for observations which might constrain the behavior of the system prior to the first ZTF detection on 2021 Feb. 4.
In the optical, ZTF's last nondetection on 2020 Dec. 21 was more constraining than those of XB-NEWS (2020 November 16, \citealp{2021ATel14415....1C}) and ATLAS (2020 December 4).

In the X-ray, no excess above a signal-to-noise ratio of three is visible in the MAXI daily count rate up to 2021 January 3 or on or after 2021 January 28; the $5\sigma$ flux limit is 0.087 photons\,cm$^{-2}$\,s$^{-1}  = 5.7 \times 10^{-9}$ erg\,cm$^{-2}$\,s$^{-1}$  (2--20\,keV) = 0.23\,Crab under MAXI's assumption of a Crab spectrum.
Similarly, \swift-BAT observations rule out at $5\sigma$ flux levels of 0.0075 counts\,cm$^{-2}$\,s$^{-1}$ (15--50\,keV) up to 2021 January 27 or starting on 2021 February 3.
Assuming a $\Gamma=2$ powerlaw with $n_H =3.1\times10^{21}$ atoms\,cm$^{-2}$ (\S \ref{sec:xray_fits}), the \swift-BAT count rate limit corresponds to a 15--50\,keV energy flux limit of $2.7\times10^{-9}$ erg\,cm$^{-2}$\,s$^{-1} = 0.15$\,Crab.
The BAT limit is thus more constraining in both flux and time.
Given the brevity of the gap in BAT coverage relative to major XRB outbursts \citep{Tetarenko:16:WATCHDOG} as well as the observed outburst duration, we consider it unlikely that an outburst above the BAT threshold was missed due to the short period of Sun constraint.
We can therefore rule out X-ray luminosities brighter than $4\times10^{37} (d/8\, \mathrm{kpc})^2$ erg\,s$^{-1}$ (MAXI) and $2\times10^{37} (d/8\,\mathrm{kpc})^2$ erg\,s$^{-1}$ (BAT) during these intervals. 
While these limits are two orders of magnitude higher than the observed \swift XRT X-ray flux (\S \ref{sec:xray_fits}), they still exclude outbursts brighter than 10\% of the peak flux of the 1999 outburst.

There is weak evidence for rebrightening or variability at late times in the outburst in the UVOT $V$ band as well as in several AAVSO measurements, although ZTF places some deeper nondetection limits at comparable time periods.

Despite the lack of strong constraints on the time of outburst onset, we can coarsely estimate the overall duration of the outburst if we assume no prior activity.
The time between the first ZTF $r$-band detection and the first nondetection is 16.9\,days; however, there was a four-night gap due to bad weather between the last ZTF detection and first nondetection.
In \swift-XRT the same calculation yields 13.2\,days, although 2.5\,days elapsed between the first ZTF detection and the first XRT observation.
Most classical XRB outbursts show a fast-rise, exponential decay temporal profile.  
If we assume the first XRT observation is the peak of the outburst, we can estimate a likely upper limit for the outburst duration by doubling the X-ray decay time.
Taking these measurements together, we estimate a plausible range of durations for this outburst at $\sim$16--26\,days, although we cannot rule out a duration of 60 days if optical activity began immediately after the last ZTF observation in 2020 December.
This duration is much shorter than normal outbursts from BH LMXBs, which typically last hundreds of days.
Less then 3\% of the outbursts cataloged in \citet{Tetarenko:16:WATCHDOG} had durations less than 30 days, and 7\% had durations less than 60 days.
However, faint X-ray transients such as this one tend to have shorter durations \citep[][and references therein]{Heinke:15:VFXRBlcs}.

\subsection{Quiescent Lightcurve} \label{sec:quiescent_lc}

\begin{figure}
	\includegraphics[width=\columnwidth]{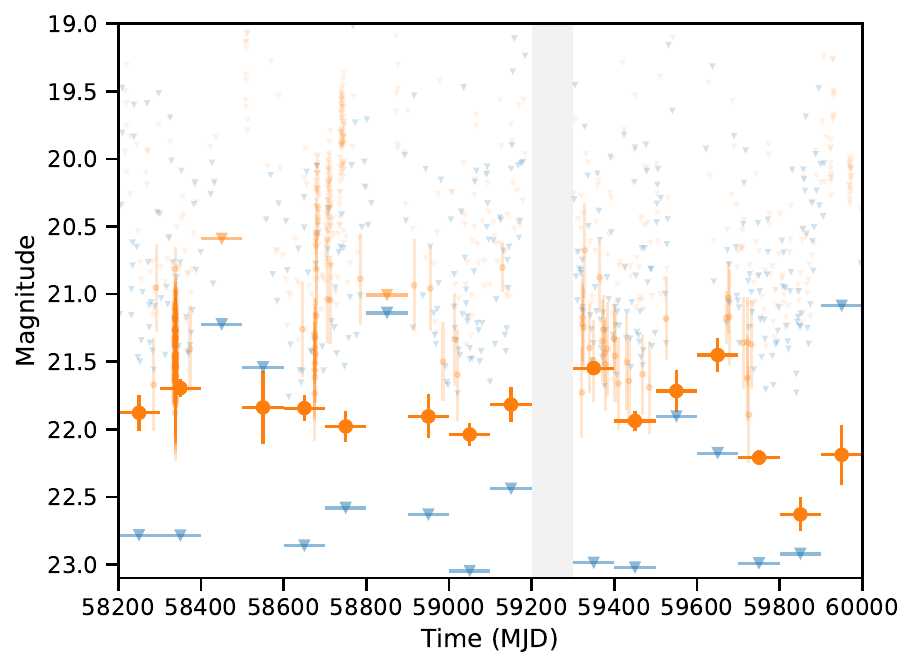}
\caption{ZTF forced photometry lightcurve of \xte in quiescence in $g$ (blue) and $r$ (orange) bands.  
We plot both single-epoch measurements (points) and 100-day error-weighted coadditions (points with horizontal error bars).
Detections with signal-to-noise greater than 3 are marked with circles, and $5\sigma$ upper limits with triangles. 
We exclude the outburst interval (gray region).
\label{fig:quiescent_lc}}
\end{figure}

Figure \ref{fig:quiescent_lc} shows the ZTF forced photometry lightcurve of \xte in quiescence from 2018 March to 2023 February.
There is clear evidence of short-timescale variability, with irregular single-epoch detections at 21.0--21.5\,mag in $r$ band.
However, averaging these measurements on 100-day timescales suggests a fairly stable average quiescent level of $m_r \sim 21.9$\,mag prior to the outburst.
\xte is undetected even in 100-day bins in $g$ band, with $5\sigma$ limits reaching $m_g \gtrsim 23.0$\,mag.

The ZTF data indicate that for at least 1000 days before and 400 days after the outburst, \xte was about three quarters of a magnitude brighter in the $r$-band than its deepest quiescent level.  
Around MJD 59800, the source faded to $m_r \sim 22.5$\,mag, near the quiescent level reported by \citet{Zurita:02:XTEJ1859Optical}.
From a detailed study of GX 339$-$4, \citet{Alabarta:21:FailedSXT} suggested that brighter O/IR emission was predictive of a failed-transition outburst, which is consistent with these observations.
However, the elevated O/IR flux levels after the outburst and the delayed fading to true quiescence appear more difficult to explain in this framework.

We also investigated the ATLAS long-term lightcurve.
Because the ATLAS data have a similar number of epochs over a longer baseline but are more than one magnitude shallower than ZTF, we did not coadd them.
The ATLAS data exhibit several high amplitude detections, including ten data points brighter than 18.5\,mag.
In all cases other than the outburst reported here, however, these detections do not form a coherent outburst temporal profile---deeper upper limits are interspersed.
These imply either fast variability on timescales of hours--days or imaging artifacts.
We examined the cutout images for these differences and found that all reported detections brighter than 18.5\,mag were due to imaging artifacts such as open-shutter readout, elevated sky backgrounds, and image differencing failures.
Accordingly we do not find evidence in the ATLAS data for other outbursts since 2015 October.

\subsection{Spectral Energy Distribution} \label{sec:sed}

As seen in Figure \ref{fig:outburst_lc}, we have few epochs with simultanous multi-wavelength coverage, making it difficult to distinguish spectral energy distribution (SED) shape from intrinsic variability.
As a point estimate, we took the P60 $g$, $r$, and $i$-band measurements obtained near the optical peak.
We dereddened these using a \citet{Fitzpatrick:99:Extinction} extinction law 
implemented in the python package \texttt{dust\_extinction} 
with $R_V = 3.1$ and $E(B-V) = 0.58$\,mag as reported in \citet{Hynes:05:XRBOIRSEDs}.
Fitting these with a blackbody using least squares, we obtain a temperature estimate of 18000$\pm$1500\,K.
As at this phase the outburst is substantially brighter than the quiescent state of the system we can be confident that the SED is dominated by the disk.
Notably, this temperature is comfortably above the $10^4$\,K temperature needed to ionize hydrogen, distinguishing this failed-transition outburst from the ``misfired outburst'' of Cen X-4 reported by \citet{Baglio:22:CenX4MisfiredOutburst}.

\subsection{X-ray Spectra} \label{sec:xray_fits}

\begin{figure}
	\includegraphics[width=\columnwidth]{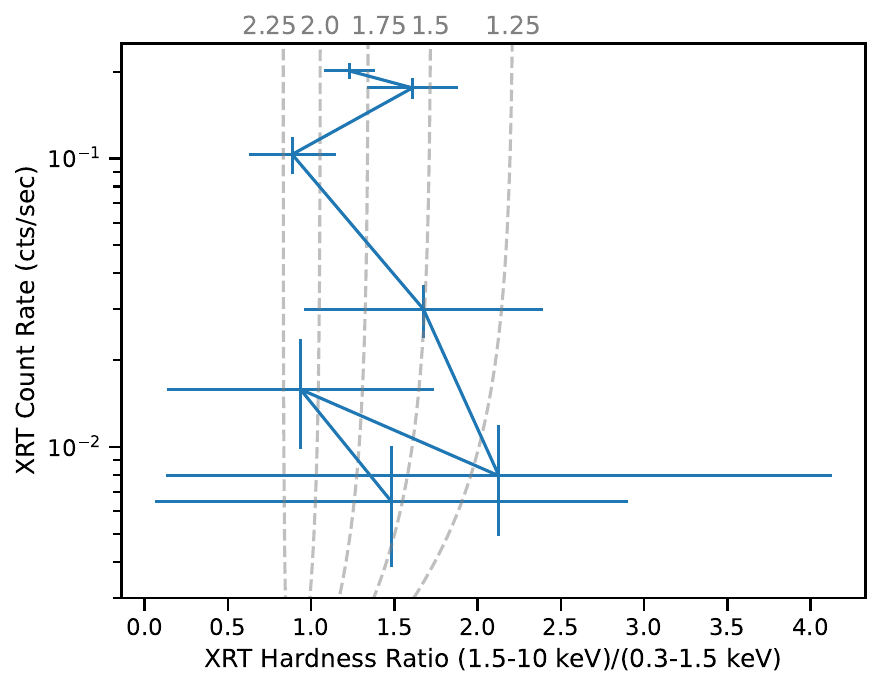}
\caption{\swift-XRT Hardness-Intensity diagram.  
Lines connect consecutive measurements; the earliest measurements are the brightest (cf. Figure \ref{fig:outburst_lc}).
Dashed lines show the hardness-intensity values produced by absorbed power law spectra with $n_H = 3.1\times10^{21}$\,atoms\,cm$^{-2}$ and spectral indices $\Gamma$ from 1.25--2.25 (top labels).
\label{fig:HID}}
\end{figure}

Figure \ref{fig:HID} shows the \swift-XRT Hardness-Intensity Diagram.
The X-ray hardness remains roughly constant as the outburst dims, remaining in a hard state throughout the observations. 

\begin{figure}
	\includegraphics[width=\columnwidth]{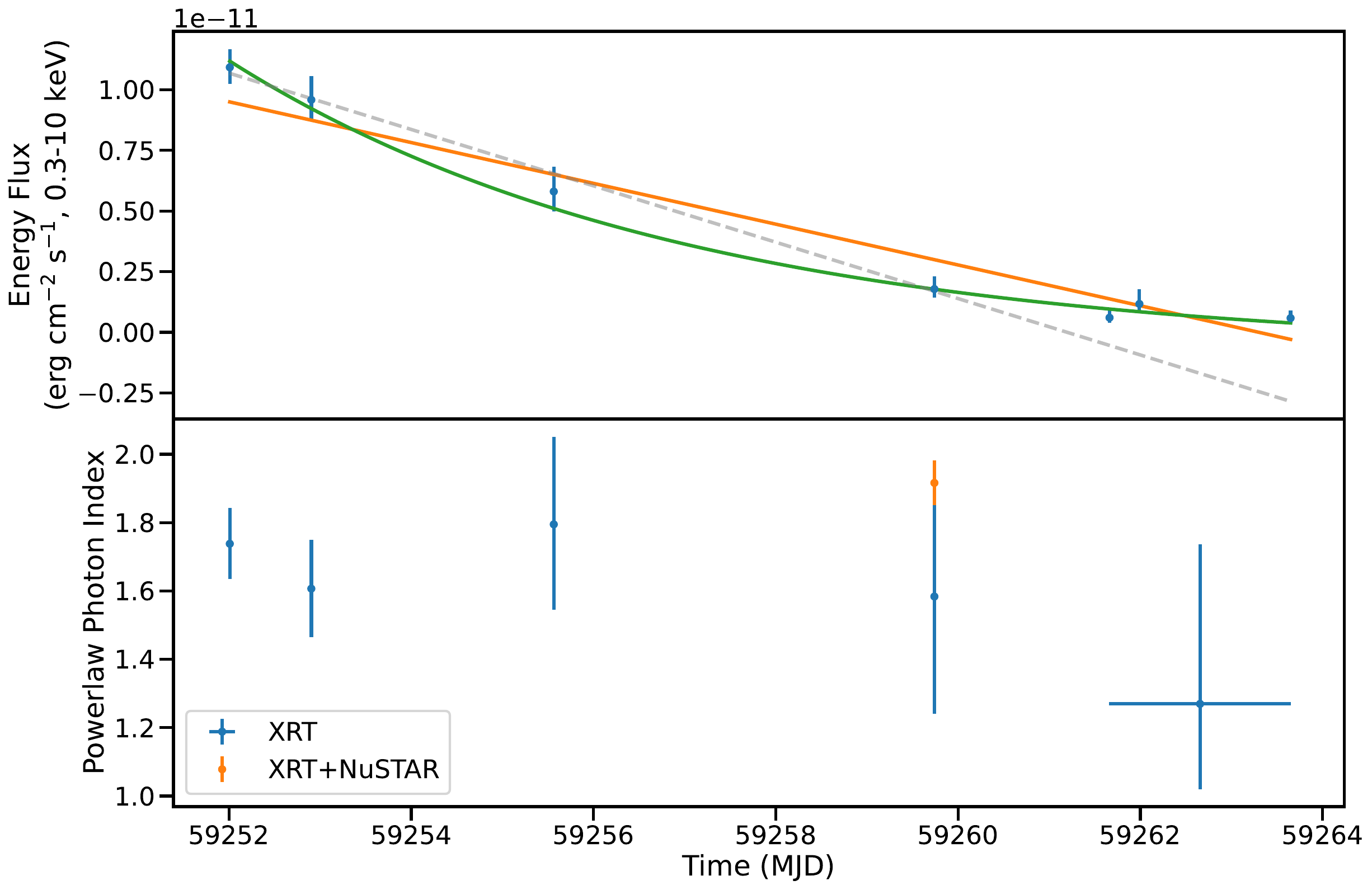}
\caption{Time evolution of absorbed power-law  parameters from spectral fits to \swift (blue) and joint \swift--\nustar (orange) data.  Top panel: 0.3--10\,keV energy flux in erg\,cm$^{-2}$\,s$^{-1}$.  The best-fit exponential (green), linear (orange), and limited-range linear (grey dashed) models are overplot.
Bottom panel: best fit spectral index $\Gamma$, where $F(E) \propto E^{-\Gamma}$.
\label{fig:spectral_fit_evolution}}
\end{figure}

\begin{figure}
	\includegraphics[width=\columnwidth]{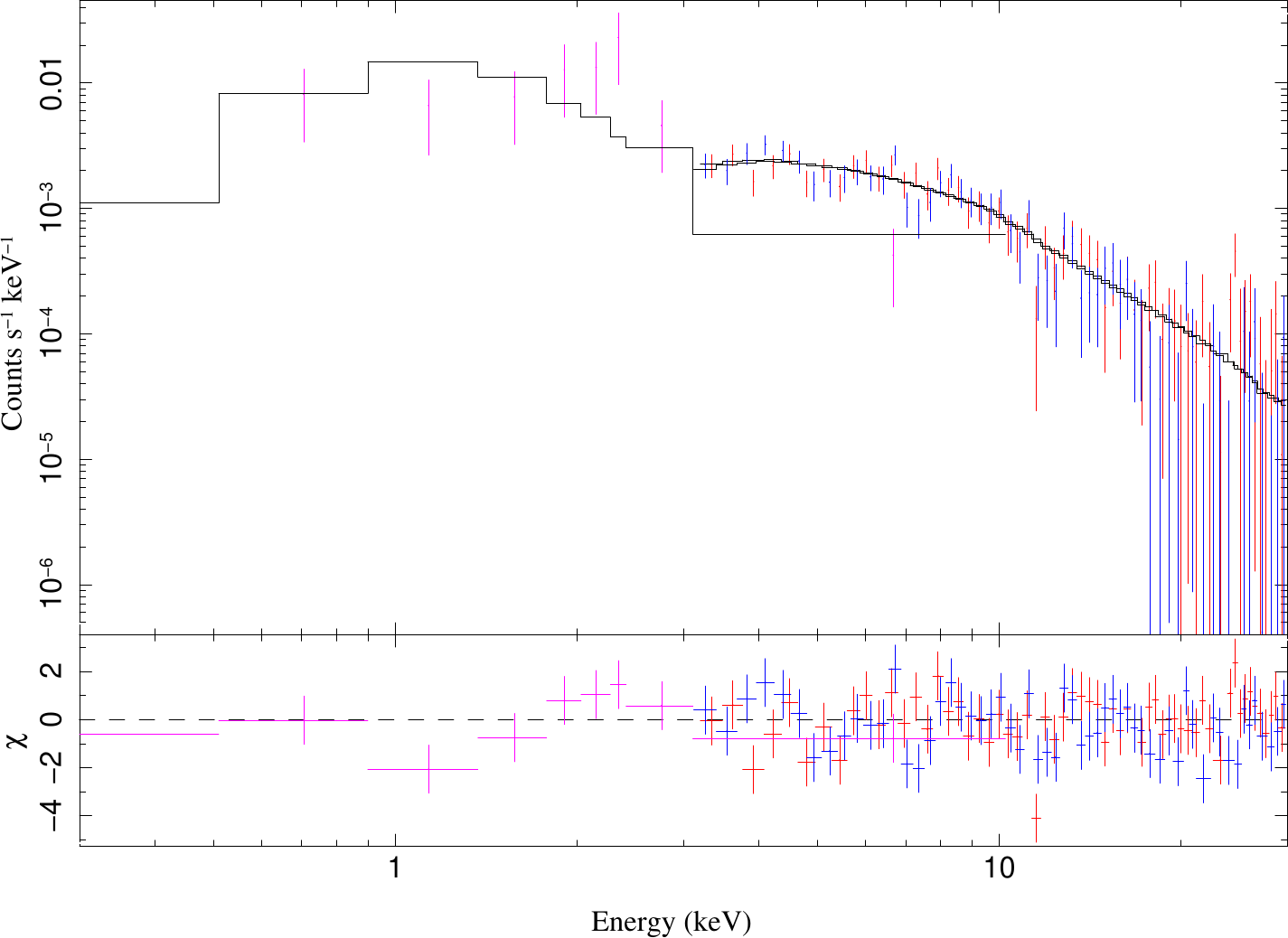}
\caption{Absorbed powerlaw fit to \swift and \nustar spectra.\label{fig:nustar_spectrum}}
\end{figure}

We performed detailed spectral fits to the epochal \swift data 
as well as joint fits of \nustar data and \swift epoch 00031827005.
The data are well fit by an absorbed power law (\texttt{tbabs $\times$ powerlaw}) in all epochs.
The powerlaw is best constrained in the joint \swift--\nustar epoch (Figure \ref{fig:nustar_spectrum}), with $\Gamma = 1.9\pm0.1$ and wstat = 129.4 for 124 degrees of freedom.
A \texttt{tbabs $\times$ diskbb} model provided a substantially worse fit, with wstat = 164.9 for 124 degrees of freedom and more systematic residuals.
We froze the column density of neutral hydrogen to its best-fit value of 3.1\spm{1.0}{0.7}$\times10^{21}$\,atoms\,cm$^{-2}$ from the first \swift epoch, as the value was more poorly constrained in the fainter later epochs.
This column density is comparable to the values reported by \citet{Farinelli:13:XTEXray} during the 1999 outburst as well as the total Galactic line-of-sight value\footnote{\url{https://www.swift.ac.uk/analysis/nhtot/index.php}} of 3.5$\times10^{21}$\,atoms\,cm$^{-2}$ \citep{Willingale:13:nH}.
The best-fit powerlaw is consistent within errors a constant value in all epochs (Figure \ref{fig:spectral_fit_evolution}), in line with the lack of evolution seen in the hardness-intensity diagram (Figure \ref{fig:HID}).

While some residual structure is apparent, we did not find strong evidence for the marginal 6.7 and 8.4\,keV emission lines reported by \citet{Draghis:21:XTENustarATel}.
As in their analysis, fits with an additional Gaussian emission line forced near these positions yielded confidence intervals for the line FWHM consistent with zero for the additional components and only modest changes in the fit statistic.  

If we assume the first \swift epoch corresponds to the maximum brightness of the outburst, we estimate a peak bolometric flux of 2.6\spm{0.7}{0.4}$\times10^{-11}$\,erg\,cm$^{-2}$\,s$^{-1}$. 
This implies a peak bolometric luminosity of $L \approx 2.0\spm{0.5}{0.3}\times10^{35} (d/8\, \mathrm{kpc})^2$\,erg\,s$^{-1}$.
The Eddington fraction at peak is thus 2.8\spm{0.7}{0.4}$\times10^{-4} (d/8\, \mathrm{kpc})^2 (M_\mathrm{BH} / 5.4\,M_\odot)^{-1}$.

\subsection{Optical Spectra}

\begin{figure}
	\includegraphics[width=\columnwidth]{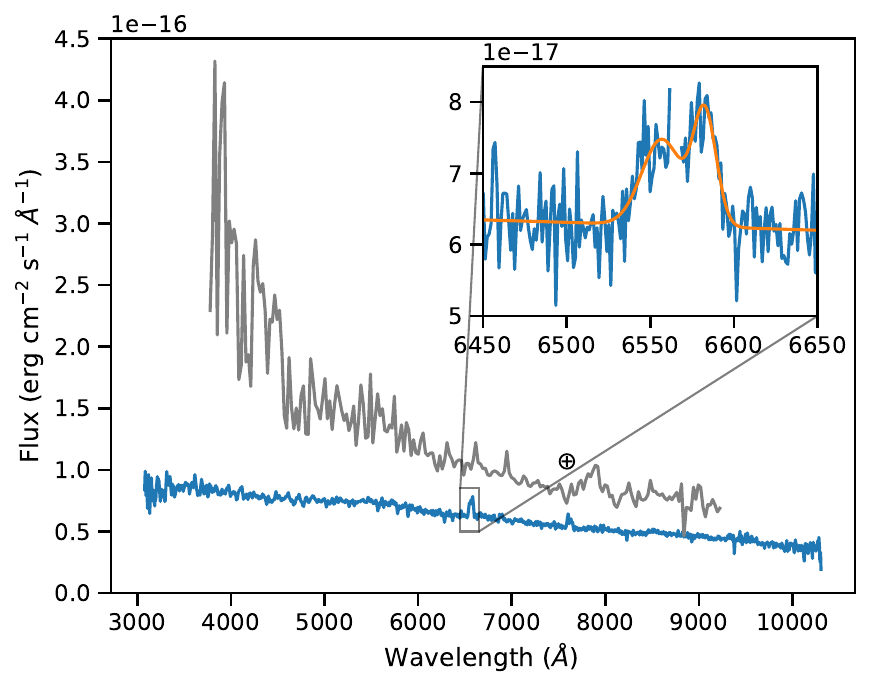}
\caption{Optical spectra (gray: SEDM; blue: LRIS) of \xte in outburst. The full LRIS spectrum is convolved with an 8-pixel boxcar filter for plot clarity.  Inset: Double-Gaussian fit (orange) to the LRIS \ha emission line.  The gap in the data corresponds to the pixels excluded due to contamination by a cosmic ray.
\label{fig:spectra}}
\end{figure}

Figure \ref{fig:spectra} shows the optical spectra of \xte at two epochs indicated by letters on the outburst lightcurve in Figure \ref{fig:outburst_lc}.
The low-resolution SEDM spectra taken near the outburst peak is noisy, with a blue continuum.
The late-time LRIS spectrum exhibits broadened \ha emission as well as interstellar absorption lines.
No other strong emission lines are present.

We fit single and double Gaussian models to the continuum-subtracted \ha line in the late-time LRIS spectrum using \texttt{pyspeckit} \citep{Ginsburg:11:PySpecKit, Ginsburg:22:Pyspeckit}.
The peak of the \ha line was unfortunately contaminated by a cosmic ray; we excluded the affected pixels.
The missing data makes it difficult to clearly identify the morphology of the line profile.
All of the model fits (single Gaussian, double Gaussian, double Gaussian with tied amplitudes and widths) were formally statistically acceptable.
Noting the asymmetry in \ha in \citet{Zurita:02:XTEJ1859Optical}, we quote and plot values and bootstrap error estimates from the asymmetric two-Gaussian fit (inset, Figure \ref{fig:spectra}), although the values derived from the other models are comparable.
The \ha line has a Full Width at Half-Maximum of 1500$\pm560$ km\,s$^{-1}$ and an equivalent width of 10.1$\pm 0.9$ \AA.
The two peaks are separated by 1170$\pm 110$ km\,s$^{-1}$.

\subsection{Disk Extent}

\citet{Casares:16:HalphaMassRatio} interpret the peak separation DP of broad \ha emission lines as the velocity of material at the outer disk radius $R_d$, with $\textrm{DP} = 2 \beta \sqrt{\frac{GM}{R_d}} \sin{i}$. 
$\beta$ is the fraction by the outer disk is sub-Keplerian; \citet{Casares:16:HalphaMassRatio} find a value of $\beta = 0.77$ is broadly consistent with their sample of BH XRBs. 
Assuming $q \lesssim 0.7 M_\odot / 5.4 M_\odot = 0.13$ for a K4V secondary of 0.7\,$M_\odot$ \citep{2000asqu.book.....C, Corral-Santana:11:XTEBH}, Equation 4 of \citet{Casares:16:HalphaMassRatio} with $\alpha=0.42, \beta=0.77$ would predict a ratio of DP/FWHM = 0.55, which is consistent within the relatively large error bars of our measured ratio of DP/FWHM = 0.78$\pm0.30$. 

Since the system masses and inclinations are constant, we can interpret changes in the peak separation as changes in the outer radius of the emitting region of the disk.
\citet{Zurita:02:XTEJ1859Optical} report two such measurements during the 1999 full outburst. 
In a first epoch near the main outburst peak, they measure 300--500\,km\,s$^{-1}$ Balmer peak separations.
In a second phase near the peak of a subsequent minioutburst, they report larger Balmer peak separations of 500--700\,km\,s$^{-1}$.
If we take the larger of each of these values and compare them to the 2021 peak separation, these imply outer disk radius ratios of 1:0.51:0.18 for the 1999 outburst peak:1999 mini-outburst:2021 outburst tail.
The corresponding optical magnitudes at epochs the spectra were obtained are about $m_R \sim$ 15.8, 18.3, and 19.5\,mag, corresponding to flux ratios of 1:0.1:0.03. 
These are broadly consistent with a naive scaling of the disk area implied by the ratio of the outer radii (1:0.26:0.03).

Estimating the absolute outer disk radius of the 2021 outburst from the LRIS peak separation, we find 
\[R_d \sim 10^{11}\,\mathrm{cm} 
\left(\frac{M_\mathrm{BH}}{5.4\,M_\odot}\right) 
\left(\frac{\beta}{0.77}\right)^2 
\left(\frac{\sin i}{\sin 70^\circ}\right)^2
\left(\frac{\mathrm{DP}}{1170\,\mathrm{km\,s}^{-1}}\right)^{-2}.\]
This is near the maximum allowed disk radius, which  
\citet{Lasota:08:UCBAccretionDisks} parameterize as
\[R_\mathrm{max} = 2.28\times10^9 f 
\left(\frac{M_\mathrm{BH}}{1\,M_\odot}\right)^{1/3}
\left(\frac{P_\mathrm{orb}}{1\,\mathrm{min}}\right)^{2/3}\,\mathrm{cm}
\]
where 
\[f = \frac{0.06}{(1+q)^{2/3}}.\]
Using the parameters of \citet{Corral-Santana:11:XTEBH}, we find $R_\mathrm{max} \sim 1.2\times10^{11}$\,cm.
This limit conflicts with our inference of a five times larger outer disk radius during the 1999 outburst, suggesting the true disk radius may be a factor of a few smaller. 
Given the uncertainties in several of the parameters, however, the overall picture remains plausible:
these data suggest that XTE J1859+226 has a large disk near its maximum allowed extent.

\subsection{Decay Profile}

Fully-irradiated accretion disks are expected to show exponential decay profiles that transition to a linear decay as the outer disk begins cooling and then to an even steeper linear decline when no irradiation is present \citep{King:98:SXTIrradiation}.
We attempted to fit an analytic exponential-to-linear temporal profile \citep{Tetarenko:18:LMXBalpha} to the X-ray flux lightcurve\footnote{The late-time XRT upper limits were not constraining so we did not include them in the temporal fits.}.
However, the observed data were insufficient to observe such a transition.  

We also fit the data with simple exponential ($f(t) = f_0 \exp(-(t - t_{\rm break})/\tau_e) + f_1$)
and linear  ($f(t) = f(t_0) (1-(t-t_0)/\tau_l)$ decays (Figure \ref{fig:spectral_fit_evolution}).
The exponential model provides a better fit to the data, with $\chi^2/\nu = 2.7/ 3 = 0.9$ compared to $\chi^2/\nu = 24.6/ 5 = 4.9$ for the linear model.
While it has two more free parameters, the exponential model is preferred by the Akaike Information Criterion \citep[AIC;][]{Akaike:74:AIC}, although the number of data points is too small for the asymptotic validity assumptions of the AIC to strictly hold.
The best-fit decay timescales were $\tau_e = 5.0\pm1.4$\,days and $\tau_l = 63.3\pm0.3$\,days.
The  exponential decay timescale is substantially shorter than those seen in full outbursts of BH XRBs, while the linear decay timescale is comparable \citep{Tetarenko:18:LMXBalpha2}.

Following \citet{King:98:SXTIrradiation}, \citet{Shahbaz:98:SXTIrradiation} identified a critical luminosity for black hole accretors $L_\mathrm{crit} = 1.7 \times 10^{37} (R_d/10^{11}\,\mathrm{cm})^2$\,erg\,s$^{-1}$. 
Below this luminosity, the outer edge of the disk will not be irradiated and the lightcurve decline should be purely linear.
Our observation of an exponential decay conflicts with this prediction:
the peak luminosity of $\sim10^{35}$\,erg\,s$^{-1}$ for this outburst (\S \ref{sec:xray_fits}) is two orders of magnitude smaller than the critical value, but the observed X-ray decline (\S \ref{sec:lightcurve}) is not linear.

In contrast, the luminosity of the 1999 outburst peaked at 2.8$\times10^{38} (d/8 \mathrm{kpc})^2$\,erg\,s$^{-1}$ \citep{Farinelli:13:XTEXray}, within a factor of two of the critical luminosity luminosity for a disk five times larger.

We consider several possibilities to resolve this apparent contradiction.
We have assumed that our first \swift-XRT measurement corresponds to the peak of the outburst.
If the outburst were already underway prior to the first \swift observations, the true peak luminosity might have been higher.
However, for our fiducial distance and disk radius, the critical luminosity almost exactly matches the limits provided by \swift-BAT monitoring (\S \ref{sec:lightcurve}).
Thus if the peak of the outburst had exceeded the critical level, it would have been detected by BAT.  
Extrapolating our best-fit exponential decay model backwards, we find that the bolometric luminosity would have been above the critical level before 2021 January 16, which is ruled out by BAT observations in that interval.

Our conversion of peak flux to luminosity uses a fiducial 8\,kpc distance to the source.
If instead we adopt the value of 14\,kpc from \citet{Corral-Santana:11:XTEBH}, the observed luminosity is a factor of three larger, still not enough to reach the critical threshold.

Similarly, a smaller disk radius $R_d$ would lower the required luminosity to ionize the entire disk.
Both $\beta$ and the inclination are poorly constrained; for a range of plausible assumed values, the disk radius could be an order of magnitude smaller.
This still leaves a gap of another order of magnitude to the critical luminosity but would allow for an unseen outburst precursor below the BAT flux limits.
Alternatively, our best-fit exponential decay timescale can be used to infer the disk radius, with $\tau_e = R_d^2/3\nu$ \citep{Shahbaz:98:SXTIrradiation}, where $\nu$ is the unknown disk viscosity.
This yields a disk radius of 
\[3.6\times10^{10} 
\left(\frac{\tau_e}{5\,\mathrm{days}}\right)^{1/2} 
\left(\frac{\nu}{10^{15}\,\mathrm{cm}^2\,\mathrm{s}^{-1}}\right)^{1/2}\,\mathrm{cm,}\]
about a factor of three smaller than the value inferred from the \ha peak separation measurement.

A final possibility to consider is that decay is not truly exponential.
While the exponential model provides an excellent fit to the data and is preferred by the AIC to a linear fit, the exponential model has four free parameters fit to seven data points.
Moreover, plateaus, reflares, and other temporal discontinuities are well-attested in the literature \citep[e.g.,][]{Chen:97:XRBLcs}.
As an ad-hoc exploration, we fit a linear decay model to the first four data points of the X-ray flux lightcurve (Figure \ref{fig:spectral_fit_evolution}, gray dashed line), which yielded $\tau_l = 61.2\pm0.5$\,days.
Since the number of data points is small the fit is reasonable, leaving the last three points to be interpreted as a plateau. 
This scenario is admittedly finely-tuned, however.

\section{Discussion} \label{sec:discussion}

Because \xte underwent a full outburst in 1999, we can be confident that the differences in the properties of this outburst are due to changes in the accretion process rather than in the fundamental system parameters.
Using the peak separation of the \ha emission lines, we inferred the outer radius of the optically-emitting disk.
While our SED fits imply that the disk was locally hot enough to ionize hydrogen, the observed peak X-ray luminosity was insufficient to ionize the entire disk by two orders of magnitude.
Surprisingly, the X-ray lightcurve showed an exponential decay characteristic of an ionized disk.
This contradicts the expectation \citep{Heinke:15:VFXRBlcs} that faint outbursts from long-period XRBs will have exclusively linear declines. 
Several strands of evidence suggests that our estimate of the outer disk radius is too high by a factor of a few, but this is insufficient to resolve the discrepancy.  
Alternatively, the observed X-ray lightcurve could be interpreted as a linear decline followed by a plateau.

The peak X-ray flux in this outburst was two orders of magnitude too low to be detected by all-sky monitors such as MAXI and \swift-BAT.
Thus without optical monitoring this outburst would have gone undetected.
However, modern synoptic surveys such as ZTF, ATLAS \citep{Tonry:18:ATLAS}, ASAS-SN \citep{Shappee:14:ASASSN}, and Gattini-IR \citep{De:20:GattiniIR} scan most of the visible sky every few nights.
When paired with an alerting system that can filter the millions of transient and variable sources they produce, these wide-field surveys can provide a powerful new approach to identifying X-ray binary outbursts despite being untargeted.
While their temporal coverage of known XRBs depends on the survey's chosen footprint and cadence, synoptic surveys can provide excellent lightcurves, as this example illustrates (cf. Figures \ref{fig:outburst_lc} and \ref{fig:quiescent_lc}).
Additionally, unlike targeted optical surveys of known XRBs, all-sky surveys can also discover brand new X-ray binaries (e.g., ASASSN-18ey/MAXI J1820+070, \citealp{Tucker:18:ASASSN-18ey}; AT2019wey, \citealp{Yao:20:AT2019wey}), particularly those at short orbital periods \citep[][and references therein]{Tetarenko:16:WATCHDOG}.
Because they are more sensitive than all-sky X-ray monitors, they provide an opportunity to identify outbursts early and to characterize samples of low-luminosity failed transition outbursts (Y.\ Wang et al., submitted).
Thanks to its depth and Southern Hemisphere site, the upcoming Legacy Survey of Space and Time conducted by the Vera C.\ Rubin Observatory \citep{Ivezic:19:LSSTScienceDrivers} will provide decade-long lightcurves for the majority of visible XRBs.
These data will provide an unprecedented real-time view of XRBs in quiescence and enable rapid multiwavelength followup of outbursts.

\acknowledgments

\facilities{PO:1.2m (Zwicky Transient Facility), PO:1.5m (SEDM),  Keck:I (LRIS), Swift, AAVSO}

\software{\texttt{FPipe} \citep{Fremling:16:FPipe},  \texttt{pysedm} \citep{Rigault:19:pysedm}, \texttt{LPipe} \citep{Perley:19:lpipe}, \texttt{PySpecKit} \citep{Ginsburg:11:PySpecKit, Ginsburg:22:Pyspeckit}, ISIS \citep{Houck:00:isis}, 
\texttt{astropy} \citep{astropy:18, astropy:2013}, 
\texttt{matplotlib} \citep{matplotlib:2007}, 
\texttt{numpy} \citep{numpy:2011, numpy:2020},
\texttt{seaborn} \citep{seaborn:2018},
\texttt{pandas} \citep{pandas:2010},
\texttt{jupyter} \citep{jupyter},
\texttt{ipython} \citep{ipython:2007}
}

Based on observations obtained with the Samuel Oschin Telescope 48-inch and the 60-inch Telescope at the Palomar Observatory as part of the Zwicky Transient Facility project. ZTF is supported by the National Science Foundation under Grants No. AST-1440341 and AST-2034437 and a collaboration including current partners Caltech, IPAC, the Weizmann Institute of Science, the Oskar Klein Center at Stockholm University, the University of Maryland, Deutsches Elektronen-Synchrotron and Humboldt University, the TANGO Consortium of Taiwan, the University of Wisconsin at Milwaukee, Trinity College Dublin, Lawrence Livermore National Laboratories, IN2P3, University of Warwick, Ruhr University Bochum, Northwestern University and former partners the University of Washington, Los Alamos National Laboratories, and Lawrence Berkeley National Laboratories. Operations are conducted by COO, IPAC, and UW. 

SED Machine is based upon work supported by the National Science Foundation under Grant No. 1106171 

The ZTF forced-photometry service was funded under the Heising-Simons Foundation grant \#12540303 (PI: Graham).

This work made use of data supplied by the UK Swift Science Data Centre at the University of Leicester.

We acknowledge with thanks the variable star observations from the AAVSO International Database contributed by observers worldwide and used in this research.

This research has made use of a collection of ISIS functions (ISISscripts) provided by ECAP/Remeis observatory and MIT (\url{http://www.sternwarte.uni-erlangen.de/isis/}). 

This research has made use of the MAXI data provided by RIKEN, JAXA and the MAXI team.

ECB, RP, and YK gratefully acknowledge support from the NSF AAG grant 1812779 and grant \#2018-0908 from the Heising-Simons Foundation.

ECB acknowledges further support from the Vera C.\ Rubin Observatory, which is supported in part by the National Science Foundation through
Cooperative Agreement 1258333 managed by the Association of Universities for Research in Astronomy
(AURA), and the Department of Energy under Contract No. DE-AC02-76SF00515 with the SLAC National
Accelerator Laboratory. Additional LSST funding comes from private donations, grants to universities,
and in-kind support from LSSTC Institutional Members.
MWC acknowledges support from the National Science Foundation with grant numbers PHY-2010970 and OAC-2117997.

\bibliographystyle{aasjournal}
\bibliography{references}

\end{document}